\documentclass{sig-alternate-05-2015}

\usepackage{amsmath}

\usepackage[tight,footnotesize]{subfigure}
\usepackage{xcolor}
\usepackage{algorithm,algorithmic}
\usepackage{amssymb}
\usepackage[noadjust]{cite}
\usepackage{flushend}
\usepackage{multirow}
\usepackage{epstopdf}
\usepackage{url}
\usepackage{gensymb}
\usepackage[pass]{geometry}

\begin{document}

\title{OWL: a Reliable Online Watcher for LTE \\ Control Channel Measurements}

\numberofauthors{1} 
\author{
\alignauthor
Nicola Bui$^{12}$, Joerg Widmer$^1$\\
$^{1}$IMDEA Networks Institute, Madrid, Spain\\
$^{2}$Universidad Carlos III de Madrid (UC3M), Madrid, Spain
}

\CopyrightYear{2016} 
\setcopyright{acmcopyright}
\conferenceinfo{AllThingsCellular'16,}{October 03-07, 2016, New York City, NY, USA}
\isbn{978-1-4503-4249-0/16/10}\acmPrice{\$15.00}
\doi{http://dx.doi.org/10.1145/2980055.2980057}

\maketitle
\hyphenation{op-tical net-works semi-conduc-tor Ah-med scar-cer speed-test syn-chro-ni-za-tion a-na-ly-sis or-tho-go-nal}

\begin{abstract}
Reliable network measurements are a fundamental component of networking research as they enable network analysis, system debugging, performance evaluation and optimization. In particular, decoding the LTE control channel would give access to the full base station traffic at a $1$ ms granularity, thus allowing for traffic profiling and accurate measurements. Although a few open-source implementations of LTE are available, they do not provide tools to reliably decoding the LTE control channel and, thus, accessing the scheduling information. In this paper, we present OWL, an Online Watcher for LTE that is able to decode all the resource blocks in more than $99$\% of the system frames, significantly outperforming existing non-commercial prior decoders. Compared to previous attempts, OWL grounds the decoding procedure on information obtained from the LTE random access mechanism. This makes it possible to run our software on inexpensive hardware coupled with almost any software defined radio capable of sampling the LTE signal with sufficient accuracy.
\end{abstract}

\ccsdesc[500]{Networks~Network measurement}
\ccsdesc[300]{Networks~Mobile networks}
\ccsdesc[100]{Networks~Network monitoring}

\printccsdesc

\keywords{Software Defined Radio, Measurements, Mobile Networks, Control Channel, LTE, Sniffer, DCI, RAR}
\section{Introduction}
\label{sec:intro}
The evolution of mobile networking technologies is continuously pushing the hardware requirements for practical experiments towards unprecedented levels and, at the same time, the academic community is required to validate novel ideas and solutions with practical tests and experiments to play an active role in the development of 5G networks. As a consequence, these experiments often require very expensive tools that are almost always the exclusive prerogative of industries and mobile operators.

To overcome this limitation a few open-source LTE implementations~\cite{nikaein2014openairinterface, demel2015lte, gomez2016srslte} provide a viable alternative to perform isolated experiments. However, for what concerns real-world network measurements current solutions either cannot provide complete and reliable information~\cite{kumar2014lte, xie2015pistream} or they demand for specific and expensive hardware~\cite{qxdm, actix, tems}.

In this paper, we introduce the Online Watcher for LTE control channel measurements (OWL). Our tool\footnote{The code is available at: \url{https://git.networks.imdea.org/nicola_bui/imdeaowl}.} is meant for researchers and SMEs that need a simple and economic solution to perform reliable measurements on LTE physical communications between mobile phones and base station. OWL is built on top of srs-LTE~\cite{gomez2016srslte} that provided us with modular and efficient implementations of LTE physical channels and basic procedures and works with a few software defined radios (SDRs), such as bladeRF~\cite{bladerf} and USRP~\cite{usrp}, capable of LTE signal sampling. 

In particular, we extended srs-LTE by implementing an online procedure to decode all Downlink Control Information (DCI) transmitted on the Physical Downlink Control CHannel (PDCCH)~\cite{lte2016channel}. Our solution is more efficient than previous attempts, because we are able to collect and maintain a list of active Radio Network Temporary Identifiers (RNTIs), which identify user equipments (UEs) within a given cell (eNodeB). In fact, RNTIs are the key for mobile phones to distinguish the control messages destined to them and to verify the success of DCI decoding.

This technique provides OWL with two very desirable features: 1) it is very reliable as it can be verified via the Cyclic Redundancy Check (CRC) field and 2) it can be executed online on inexpensive hardware, since it does not need heavy computation. 

We measure OWL's reliability by comparing the schedule information obtained from DCIs to the used network resources by means of power measurements on the raw signal: in more than $99$\% of the captured frames in our tests OWL detects all the scheduled transmission, scoring an average $99.85$\% successful decoding ratio overall. Therefore, OWL can be used as a ground truth check for mobile phone measurements, to perform extensive mobile networks measurement campaigns or to evaluate mobile networks performance and functionalities.

The rest of the paper consists of a comparative review of the related works in Section~\ref{sec:related}, the basic LTE details to understand how OWL works in Section~\ref{sec:control}, the description of OWL and its architecture in Section~\ref{sec:architecture}, OWL's performance evaluation and an example of measurements obtained with it in Section~\ref{sec:results} and our conclusions in Section~\ref{sec:conclusions}.
 
\section{Related work}
\label{sec:related}
To the best of our knowledge the first non-commercial attempt to decode LTE control information has been LTEye~\cite{kumar2014lte}: DCI messages are not encrypted, but only the intended receiver can verify the successful decoding, because the CRC field of the message is scrambled (binary exclusive OR operation) with the UE's RNTI. To decode DCIs without knowledge of the destination RNTI, LTEye, first, assumes the decoding to be successful, then obtain the destination RNTI from the CRC field of the message XORed with the CRC computed on the decoded data. The shortcoming of this is that the CRC cannot to be used to validate the decoding operation. To solve this, the authors propose to re-encode the decoded message and to compare the result with the original bits received before the decoding operation. Although feasible under almost bit-perfect channel condition, this approach suffers from low reliability as has been verified in~\cite{xie2015pistream}. 

The latter paper proposes RMon, another technique to monitor the resource allocation on the Physical Downlink Shared CHannel (PDSCH): by comparing the received signal strength to LTE reference signals~\cite{lte2016phy}, RMon is able to evaluate which resource block is used regardless of the control information. Although quite reliable, this approach does not allow to obtain any additional information beyond the fraction of used resources.

Instead, thanks to the list of active RNTIs, OWL is both reliable, because it can verify the DCI decoding with the CRC, and expressive, since it can access all DCI fields. Of course, commercial products might offer similar features albeit at a much higher price and complexity, e.g., QXDM~\cite{qxdm}, Actix Analyzer~\cite{actix}, or TEMS investigation~\cite{tems}.

For what concerns open-source LTE implementations, we use srs-LTE~\cite{gomez2016srslte} for its very efficient implementation. In addition, the modularity of the architecture and the adherence to the standard terminology allowed us to realize OWL starting from the provided example program to record and synchronize the LTE signal. Alternative approaches include gr-LTE~\cite{demel2015lte} a solution based on GNU Radio, and openLTE~\cite{nikaein2014openairinterface}, which is more focused on the actual transmission and reception of PDSCH and is more suitable for isolated experiments where both UEs and eNodeB are controllable. 

\section{Control Channel Decoding}
\label{sec:control}
This section is a mini-guide to LTE physical channels and procedures needed to understand the operations performed during the control channel decoding. In particular, we cover synchronization procedures and the related channels, RNTI types and the random access procedure and, finally, the control channel and the information carried by DCI messages. In what follows, we limit our description to frequency-division duplex and standard cyclic-prefix duration and most of LTE's subtleties are omitted due to size limitation of the paper. The interested reader is referred to~\cite{lte2016phy} for other details of LTE.

\begin{figure}[t!]
\centering
\includegraphics[width=\columnwidth]{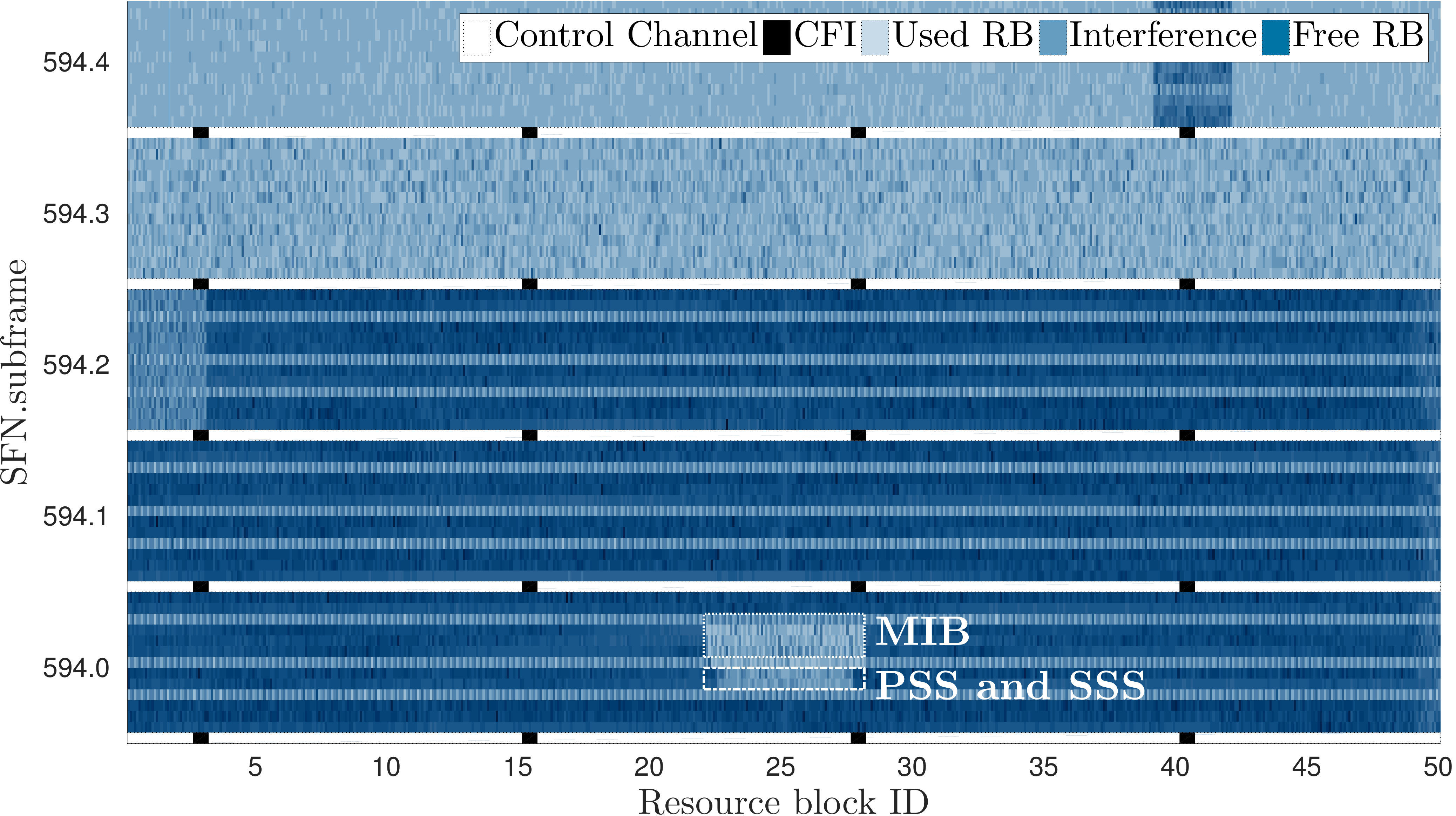}
\caption{Annotated capture of half a frame of a $10$ MHz LTE signal. The OFDM grid spans resource blocks on the $x$-axis and subframes on the $y$-axis. We highlighted the synchronization sequences (PSS and SSS) and the MIB in the center of the band. The horizontal lines representing the control channel are highlighted in white, while the CFI elements within the control channel are drawn in black.}
\label{fig:legen_noa}
\end{figure}

Figure~\ref{fig:legen_noa} is an annotated power chart of half a frame of a $10$ MHz LTE signal. It is obtained by expanding the orthogonal frequency-division multiplexing (OFDM) grid in $600$ sub-carriers ($x$-axis) and $70$ symbols ($y$-axis). A resource element (RE) is the minimum two-dimensional unit ($1$ sub-carrier $\times$ $1$ symbol), a resource block (RB) consists of $84$ REs organized over $7$ symbols and $12$ subcarriers, $7$ symbols form a slot, $2$ slots form a subframe and $10$ subframes are a $10$ ms frame. 

In order to synchronize with a given eNodeB, the user equipment (UE) computes the correlation between the received signal and three known Zadoff-Chu sequences. This allows the UE to acquire the location of the Primary Synchronization Sequence (PSS) and to decode the Secondary Synchronization Sequence (SSS). Both can be found in subframes $0$ and $5$ in every frame. By doing so, the UE can compute the eNodeB Physical Cell ID (PCI) and the system timing, which are needed to identify all the remaining physical channels in LTE. The next synchronization step is decoding the Master Information Block, which is located in subframe $0$ of every frame and carries the System Frame Number (SFN) as well as other system parameters.

RNTIs are $16$-bit identifiers used by the eNodeB to distinguish among the many UEs connected at any given time. Among the different types of RNTI, only two are relevant to our procedures: random access RNTI (RA-RNTI) and cell RNTI (C-RNTI). The former only takes values in $[1-10]$ and is used during the random access procedure to allow the eNodeB to address an unknown UE. The latter can take any unreserved value in $[0x003{\rm D}-{\rm FFF}3]$ and is assigned to the eNodeB at the end of the random access procedure. 

A brief overview of the random access procedure is as follows: 1) the UE sends one out of 64 possible preambles (Zadoff-Chu sequences) in subframe $i$; 2) the eNodeB sends a Random Access Response (RAR) message in which a temporary C-RNTI is assigned to the UE; 3) the UE sends a RRC connection request message; 4) the eNodeB responds with contention resolution message to UE. In order for the UE to receive the RAR, the related DCI is sent to the RA-RNTI address $i+1$, which is defined by the subframe where the UE sent the preamble. The C-RNTI received during step 2 is only confirmed in step 4; in fact, if two or more UEs selects the same subframe for sending the preamble, all of them receives the RAR with the same information. However, only one of them will successfully complete step 3, thus, receiving the final confirmation from the eNodeB. In any case, the temporary C-RNTI sent in the RAR is assigned to one of the users participating in the random access procedure. 

Note that the DCI sent to the RA-RNTI only carries information for the UE to decode the RAR, but the actual RAR is a proper RRC message sent in the shared downlink channel. Thus, the UE can decode the 6 bytes of the actual message, which consists of a short header, the time alignment, the upload grant to let the UE send the message in step 3 and, in the last 2 bytes, the C-RNTI that is going to be used by the user winning the contention.

LTE schedule is completely governed by the eNodeB and no communication can happen without an explicit control message being issued on the control channel that occupies the first symbol(s) of each subframe. In the figure, we colored all control channel symbols in white for an easier identification, whereas the remaining REs are colored in different shades of blue (light, dark and medium for used, free and interfering RBs). The actual number of symbols used for the control channel is specified in the Control Format Indicator (CFI) a $32$-bit sequence spanning $16$ RE, the position of which depends on the PCI (in black within the control channel in Figure~\ref{fig:legen_noa}) and that can assume a value in $\{1,2,3\}$. Depending on the size of the control channel and the system bandwidth, UEs need to monitor different locations on the control channel, since, to avoid collisions, a control message destined to a given RNTI can only occupy a subset of the available locations.

Due to LTE's flexibility and its many revisions, there exist many different DCI formats. However, here we only provide the common characteristics that allows OWL to monitor the cell traffic. First of all, every DCI format specifies whether it is related to the uplink or the downlink: this information is either derived by the size of the message, if it is unique for a given format, or by the first bit of the message, otherwise. The second field which is always present in transmission related DCIs is the Modulation and Coding Scheme (MCS) field: $5$ bits that specify the modulation and the code rate that will be used in the corresponding transmission. The last two pieces of information that OWL extracts from DCIs are the number of used resource blocks $N_{RB}$ and the transport block size. The definition of the former depends on the actual DCI format, while the latter is derived by using MCS and $N_{RB}$ as indices in a lookup table. The complete definitions can be found in~\cite{lte2016proc}. Finally, DCI messages have a CRC footer, which is the result of a XOR operation between the actual CRC computed over the message payload and the C-RNTI of the destination UE.

\vspace{.2cm}
\section{OWL Architecture}
\label{sec:architecture}

The OWL software architecture is composed of three processes: 1) a \emph{synchronized signal recorder}, 2) the actual \emph{control channel decoder}, and 3) a \emph{fine-tuner} that is used when a control message is expected to be found on the control channel, but the main process cannot decode it. Finally, we develop an auxiliary \emph{verifier} tool that checks whether the decoded DCIs match the actual resource allocation on the PDSCH.

\subsection{Synchronized signal recorder}
\label{sec:arch:rec}

OWL's signal recorder inherits most of its functionalities from the synchronized signal recorder provided by srs-LTE. This tool, first, synchronizes the software to the eNodeB transmissions by means of PSS and SSS correlation, then acquires the remaining information by decoding the MIB, and finally it writes an output file starting from the first symbol of the first frames for which it obtained a successful MIB decoding. However, it might happen that the system synchronization degrades without the recorder being able to notice, in particular for recordings longer than a few seconds.

To improve this, we provide the recorder with a synchronization check at the beginning of every frame. In addition, we provide the recorded trace with an error log that tracks any synchronization issues and any other software related error that might hamper the following operations.

\subsection{Control channel decoder}
\label{sec:arch:dec}

OWL's main component is the control channel decoder. It can work either online while the signal is being sampled by the SDR or offline processing prerecorded traces. Our control channel decoder inherits from srs-LTE the basic decoding functions, such as CFI decoding, channel equalization and mapping. However, srs-LTE provides all the functionalities as they would have been implemented in a UE. Instead, OWL needs these functions to be extended to cover all possible control channel allocation: in particular, while a single UE can monitor a limited set of control channel locations, OWL needs to extend the procedure to all possible locations and DCI formats.

In any case, both srs-LTE and OWL only perform actual DCI decoding if there is an ongoing transmission on the REs of the scanned location. If this is the case, the decoding procedure is repeated for all possible DCI sizes. srs-LTE considers the decoding operation successful if the CRC field, scrambled with the CRC computed on the data, matches the C-RNTI of the UE under test. Instead, OWL only requires that any of the C-RNTI of the active list matches with the decoded message.

Since the C-RNTI list is empty when the system starts, OWL needs to populate it while decoding the control channel. To do so, OWL can either 1) exploit the random access procedure or 2) verify the decoding success by re-encoding the DCI as LTEye does. In the former procedure, whenever a DCI is decoded with the CRC field XORed with a RA-RNTI ($[1-10]$), not only is it considered a successful decoding, but also the RAR message, which is sent in PDSCH at the RBs specified in the DCI by means of MCS and $N_{RB}$, is actually demodulated and decoded and provides OWL with a new C-RNTI to be inserted in the active list. 

LTE RRC messages are coded using ASN.1~\cite{lte2016rrc}, but the particular configuration of the RAR messages allow us to simplify the decoding by just taking the last two bytes of the message, because the C-RNTI is always specified in this location. In addition, since the actual RAR message is provided with a CRC field, OWL is able to evaluate the correctness of the whole operation by verifying the message checksum against the CRC field.

Also, OWL implements LTEye re-encoding procedure to bootstrap the list for those C-RNTIs that were assigned before the logging started and to recover from the missed random access procedures in the unlikely event of de-synchroni-zation. This gives us the added benefit to be able to compare the two methodologies: whenever a transmission is detected on the control channel we verify it both by checking the re-encoded message against the received symbols and by checking whether the C-RNTI is in the active list.

C-RNTIs are just temporary identifiers and, after a complete SFN cycle ($10.24$ seconds) of inactivity, a UE needs to perform the access procedure again to obtain a new one. For this reason, OWL resets all the RNTIs in the list that are inactive for more than a SFN cycle. Finally, while OWL uses the LTEye re-encoding procedure to bootstrap the RNTI list, at steady state we verified that OWL effectively detects all new RNTIs assigned by the eNodeB. As such, we only enable the DCI re-encoding when OWL detects a DCI message whose CRC is not XORed with a C-RNTI in the active list. This makes OWL both robust, because of the actual decoding verification, and computationally effective, because unneeded re-encoding operations are avoided.

We thoroughly evaluate the offline control channel decoder performance and, on a single Core i3 processor, the overall computational time is about half the length of the recorded trace. Similarly, the online decoder works without ever interfering with data stream arriving from the SDR.

\subsection{Fine-tuner}
\label{sec:arch:tun}

While, theoretically, the control channel decoder should be able to decode all DCIs, we identify a few rare conditions for which power is detected on the control channel, but no DCI message has been decoded. We believe that these conditions are due to either equalization or synchronization problems. The fine-tuner is able to correct the majority of these issues by iteratively performing the decoding operation on the specific location only and varying the timing offset of the LTE signal.

The drawback of the fine-tuner is that its operation time is proportional to the number of uncertain control channel locations. Our tests show that the fine-tuner takes less than the trace duration in $50$\% of the cases, less than five times the trace duration in $90$\% and up to ten times in the remaining $10$\%. However, they also show that the fraction of DCI message fixed by the fine-tuner is always lower than $5$\% of the overall decoded messages and the actual fraction is independent of the time taken to decode it; in fact the time only depends on the number of uncertain locations.

\subsection{Pipeline}
\label{sec:arch:pip}

\begin{figure}[t!]
\centering
\includegraphics[width=1\columnwidth]{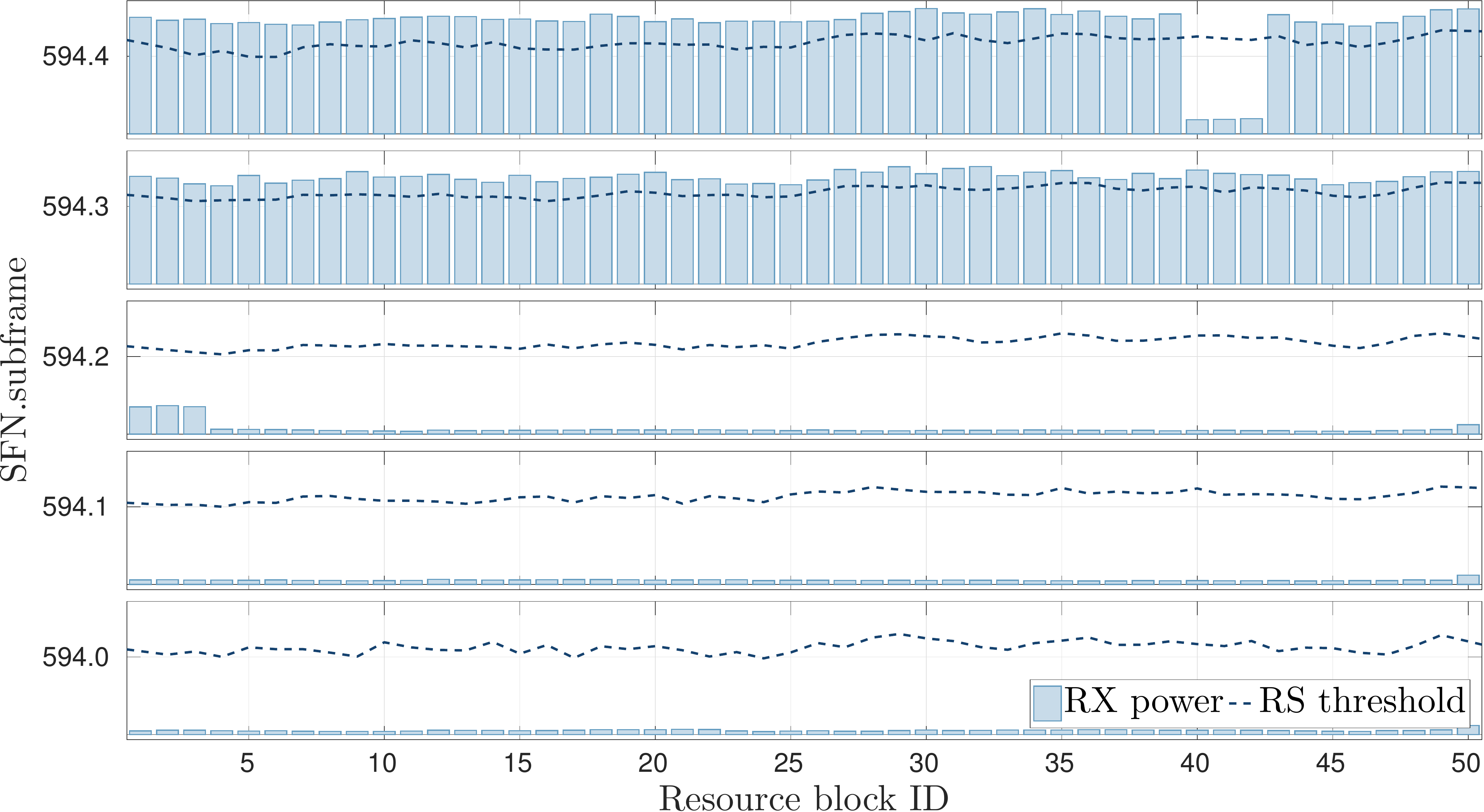}
\caption{Results of the verifier on the same locations of Figure~\ref{fig:legen_noa}: the five rows compare the reference signal threshold to the average power measured on the RBs.}
\label{fig:power}
\end{figure}

\begin{figure*}[t!]
\centering
\subfigure{
\includegraphics[height=.49\columnwidth]{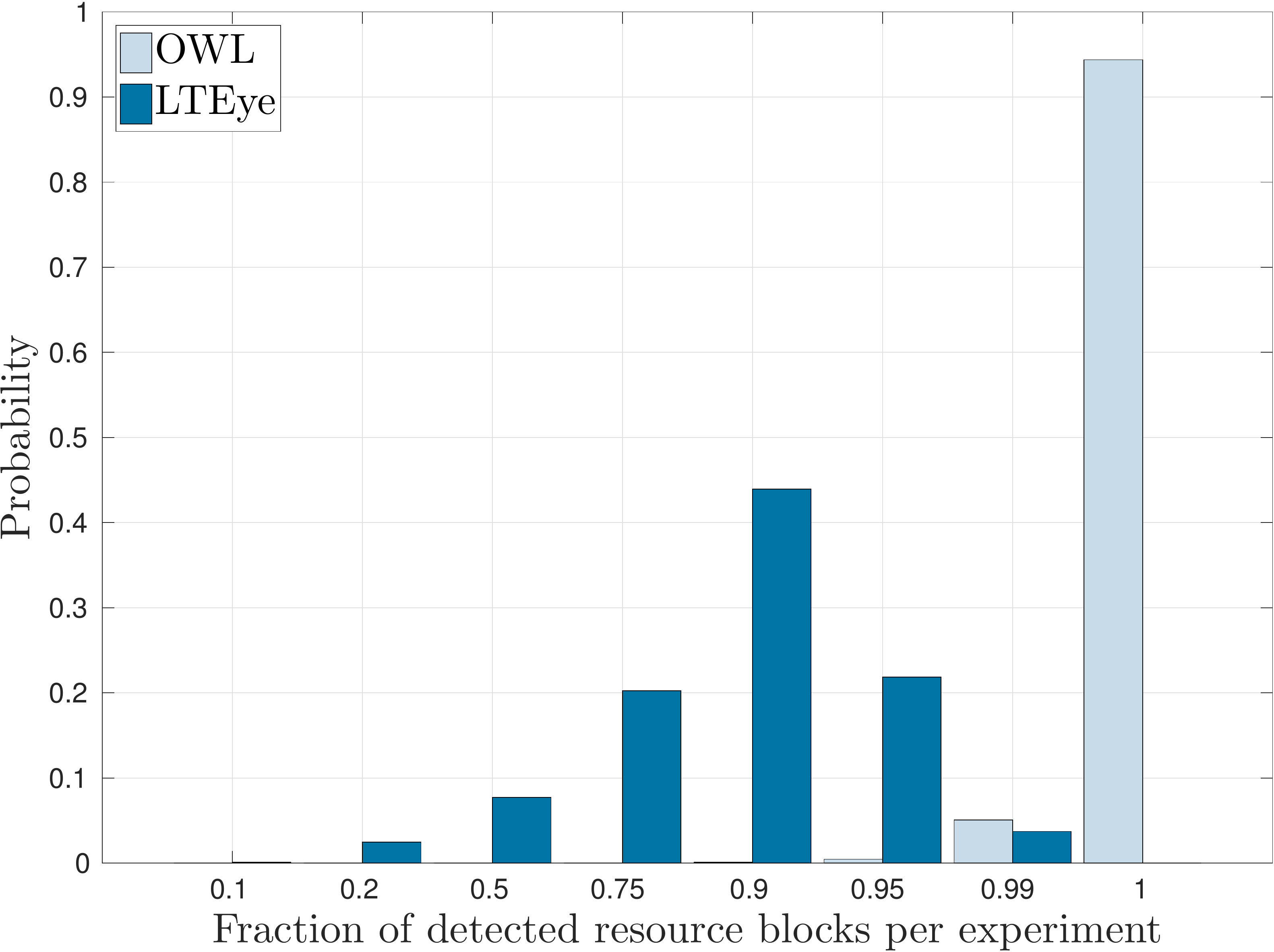}
\label{fig:epe}}
\subfigure{
\includegraphics[height=.49\columnwidth]{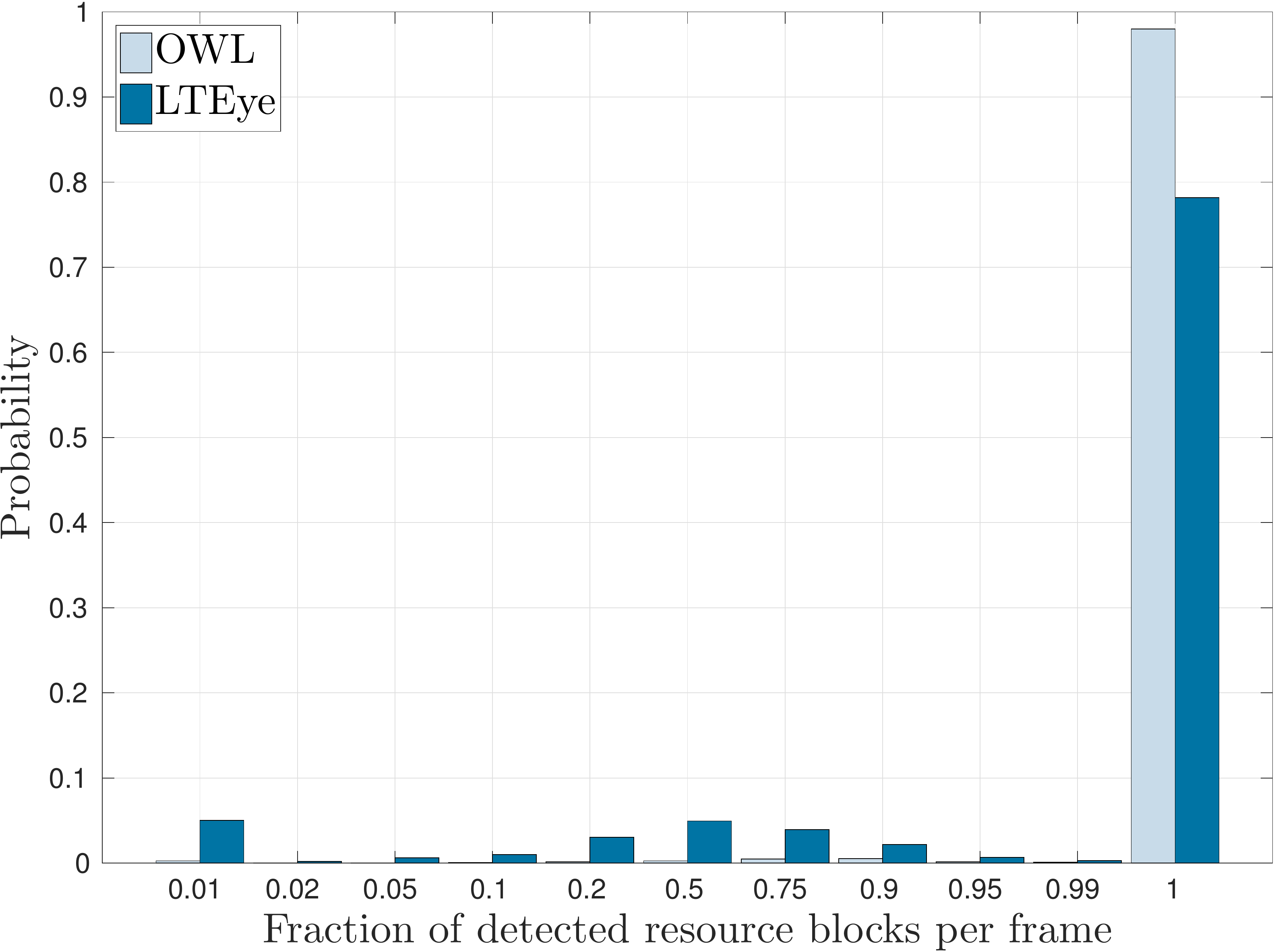}
\label{fig:epf}}
\subfigure{
\includegraphics[height=.49\columnwidth]{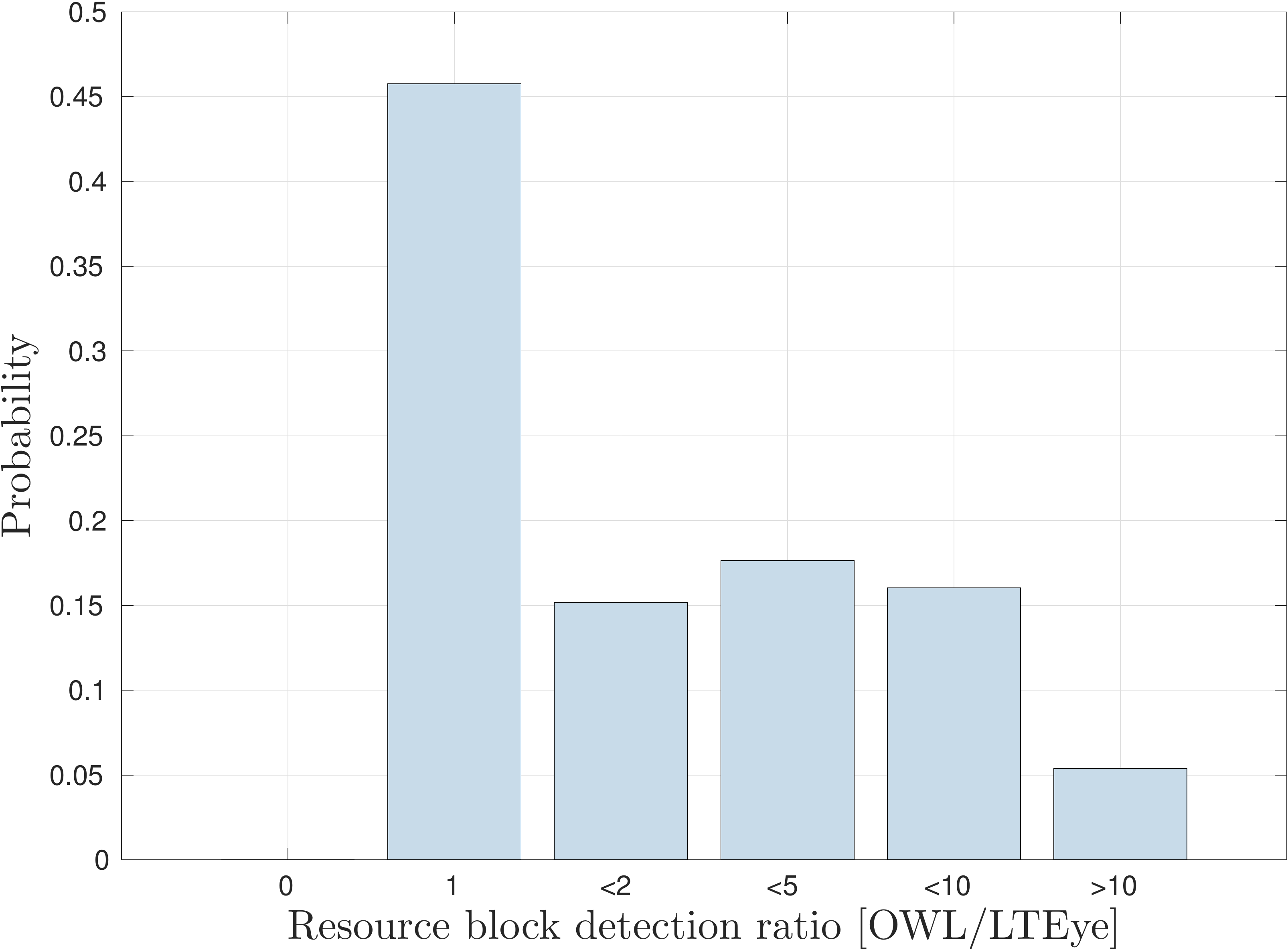}
\label{fig:gpf}}
\caption{OWL validation campaign results. On the left plot we show the likelihood ($y$-axis) of detecting the fraction of RBs specified on the $x$-axis computed per frame; the central plot is similar, but measured per frame; the plot on the right illustrates the ratio between the number of decoded RBs of OWL and LTEye.}
\label{fig:compare}
\end{figure*}

The overall OWL solution coordinates as many parallel processes as cores are available on the CPU denoted by $k$. The first process continuously records the LTE signal and cyclically switches the saving location among $k$ files. These files are located on ramdisks in order not to interfere with the trace recording itself. As soon as the first process switches to the next saving location, the second process runs the control channel decoder on the recorded trace. This process produces the main output and identifies whether and where there are uncertain locations in the control channel trace. As soon as the second process is done, a new process is started to run the fine-tuner on the trace. In case the fine-tuner takes longer than the time by which the first process needs again the file to save the next trace, we force the fine-tuner to timeout before this happens. In this way OWL might lose a few control messages, but does not stop the trace recording. Also, the fine-tuner processing can last at least $k-2$ times the length of the recorded trace and OWL hardware can be chosen to reduce the likelihood for this to happen to a minimum.

\subsection{Verifier}
\label{sec:arch:ver}

Finally, to verify whether the decoded information matches the actual PDSCH resource allocation, we develop a simple tool that takes as inputs the decoding log and the raw LTE signal trace. For each subframes it computes how many RBs are detected by OWL by summing all the $N_{RB}$ values of downlink messages. Similarly, it evaluates for each subframe and for each RB whether the average power measured on the PDSCH is higher or lower than the power measured on the reference signals that are the closest to the related RB. While the control channel decoder can decode both uplink and downlink schedule, the verifier tool can only measure downlink information with a single SDR, because in FDD systems the uplink physical channel is separated from the downlink by a few hundred MHz. Hence, in this paper we can only systematically verify the downlink schedule and we leave the development of a verifier tool for the uplink channel using two SDR for future work.

Figure~\ref{fig:power} visualizes the result of the power analysis of the verifier tool performed on the same frame used in Figure~\ref{fig:legen_noa}. By comparing the two figures, it can be seen that the power analysis can effectively identify ongoing transmission (lighter areas of Figure~\ref{fig:legen_noa} correspond to taller bars in Figure~\ref{fig:power}). Also, the first RBs of subframe 594.2 are correctly identified as interference.

\subsection{OWL release details}
\label{sec:arch:ver}

OWL extends srs-LTE by adding the following:
\begin{itemize}
 \item support for DCI formats 1B, 1C, 1D, 2, 2A
 \item automatic decode of DCIs sent to RA-RNTIs
 \item random access response messages decoding
 \item C-RNTI list management
 \item DCI verification by re-encoding. 
\end{itemize}

In order to run OWL, the SDR must support a LTE compatible sampling rate: $30.72$ Msps (samples per second) to be standard compliant, but we successfully tested OWL at $23.04$ Msps for $20$ MHz bandwidth and $11.52$ Msps for $10$ MHz. The PC must be able to receive the recorded stream from the SDR and store it: this can be achieved by means of USB3, 1Gbit Ethernet (up to $10$ MHz only) and 10Gbit Ethernet; although we did not try less capable devices, we successfully used OWL on Core i3 PCs by temporarily storing and decoding the traces in RAM and only using the physical disk to log DCI information.

At the moment of writing this paper, OWL's alpha release is already available at \url{https://git.networks.imdea.org/nicola_bui/imdeaowl} and is being tested by a small group of colleagues. We currently plan to run the alpha testing until the end of September and to release a fully documented beta version by the conference date in the same repository. OWL is completely open-source and it is released under the Affero General Public License v3.

\section{Results}
\label{sec:results}

In this section we provide two sets of results: a first set validates OWL and compares it to LTEye, while second provides an example of analysis realized with our tool. All the tests of this section are performed by capturing a $10$ MHz LTE channel in the frequency band at $1854.1$ MHz through a BladeRF x40 SDR connected to $4$-processor Core i3 mini PC equipped with $4$ GB of RAM and running Ubuntu 14.04.

In order to compare OWL and LTEye we run more than a thousand experiments in which we recorded $5$-second traces that we subsequently decoded with OWL. In all experiments we let the fine-tuner process end, to obtain the maximum number of decoded messages. 

To evaluate the performance of LTEye, after each DCI decoding we verified its success by re-encoding the message and comparing it to the received signal. If the two differ for less then $2$\% of the bits we count the message as a valid decoding for LTEye. Note that, for the sake of fairness we compute this after having processed the trace with the fine-tuner in order to compare OWL's procedure based on random access to LTEye re-encoding solution. Also, we choose the test location in order to have the best possible reception in our space from a nearby eNodeB.

Finally, we compute the number of RBs effectively used in PDSCH by running the verifier tool on the raw captures. Figure~\ref{fig:compare}~(left) evaluates the fraction of RBs detected by OWL and LTEye compared to those detected by the verifier in each frame. We group the results in bars that show in the ordinate the probability to successfully decode a given fraction of RBs ($x$-axis) for the two solutions. In all figures the $x$-axis is modified in order to highlight where the probability distributions concentrate. OWL decodes all the RBs in about $95$\% of the tests and in the remaining $5$\% only miss $1$\% of them. Conversely, the most frequent result for LTEye is decoding $90$\% of the RBs and it never successfully decode all RBs in a test.

The central plot shows a similar result, but, instead of evaluating the detection ratio averaged over experiments, it plots the results for each frame: OWL successfully decode all RBs in almost $99$\% of the frames, while LTEye achieves less than $80$\%. Both solutions have non-zero probability for all detection ratios, since different frames might have a variable number of allocated RBs and a single error may represent different detection ratios depending on the actual load. 

The last plot of the first set shows the ratio between the detection ratio of OWL and that of LTEye: the two solutions decodes the same number of RB in $45$\% of the frames ($x=1$), OWL never decodes less RBs than LTEye ($x=0$), but consistently decodes a larger number of RBs in the majority of the frames, providing an improvement larger than a factor of $10$ in $5$\% of the frames and an average improvement larger than a factor of $4$.

Although mis-detection happens with higher probability, both solutions can also generate false positives, for instance due to strong noise/interference. However, since in our tests false positives have been detected in very few tests only, we deem their impact negligible. 

\begin{figure*}[t!]
\centering
\subfigure{
\includegraphics[height=.49\columnwidth]{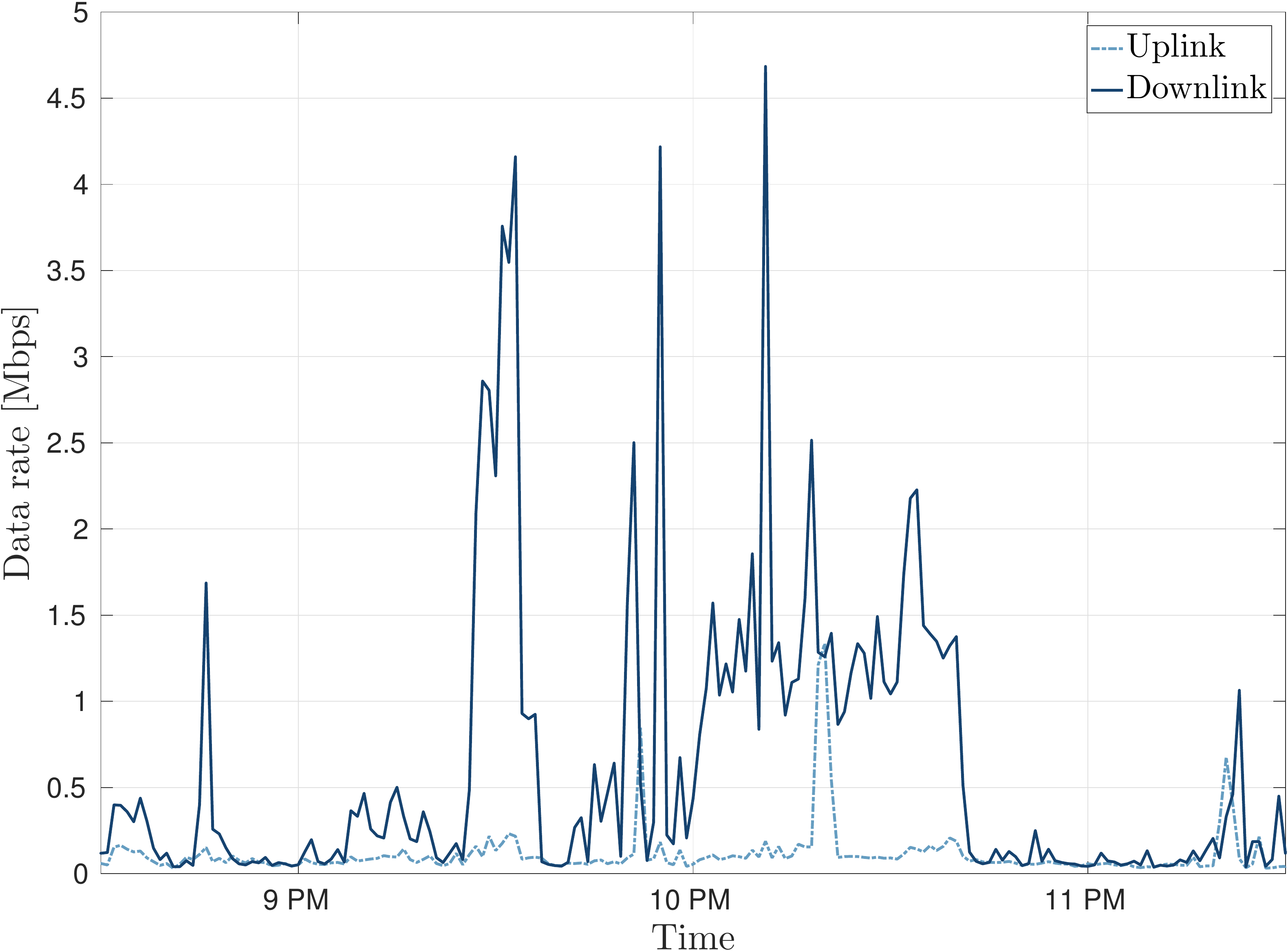}
\label{fig:day}}
\hspace{.1cm}
\subfigure{
\includegraphics[height=.49\columnwidth]{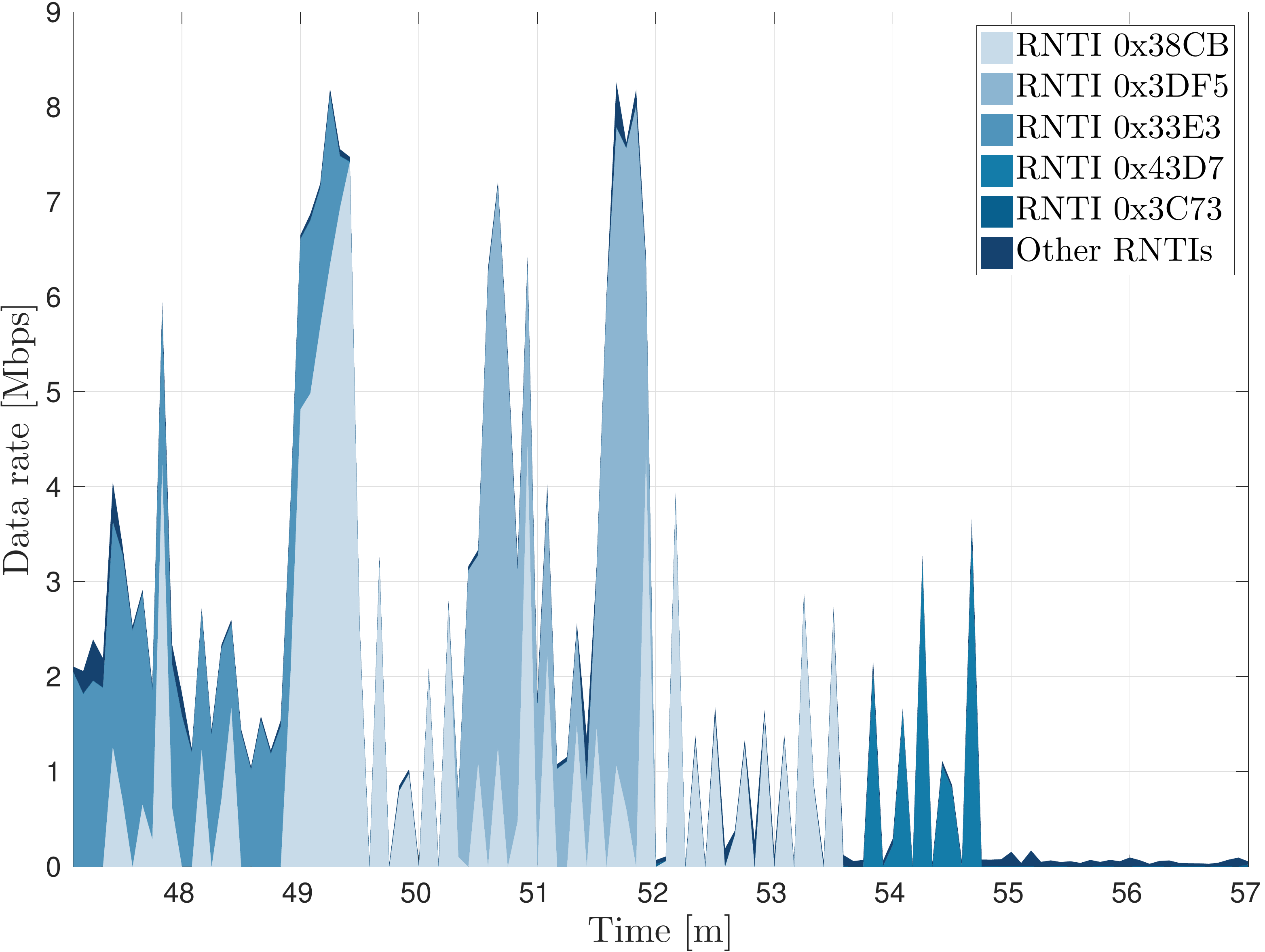}
\label{fig:hour}}
\subfigure{
\includegraphics[height=.49\columnwidth]{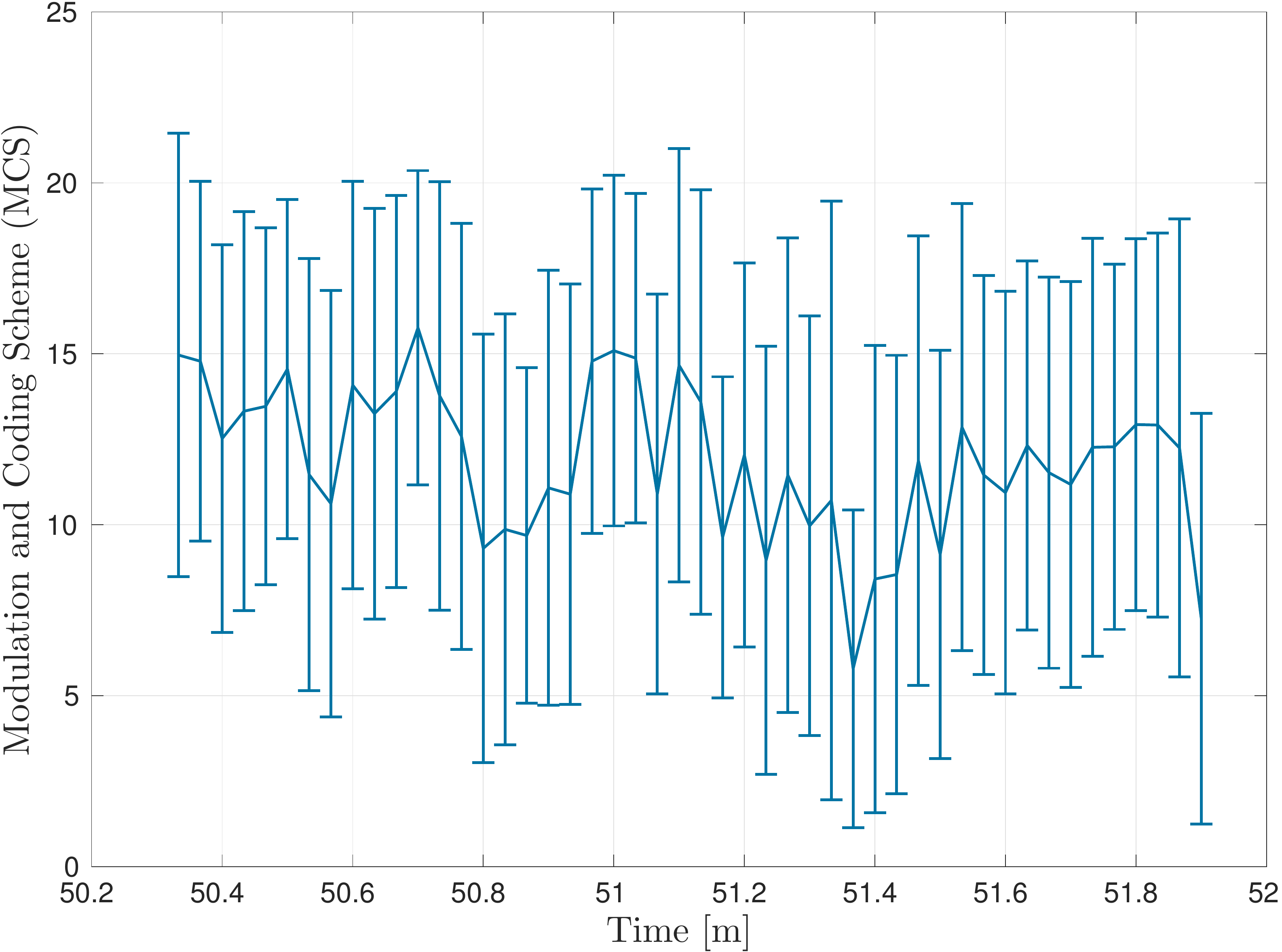}
\label{fig:cms}}
\caption{Second set of results: among the full day measurement we plot the total data rate during $3$ hours (left), a zoom of $10$ minutes showing individual users (center) and a plot of about $2$ minutes of the second user's MCS variation (right).}
\label{fig:example}
\end{figure*}

To conclude the results section, Figure~\ref{fig:example} shows a set of graphs obtained from a full day measurement campaign: the figure on the left shows the overall downlink and uplink data rates in a $3$-hour window starting at 8:30 PM averaging the results over $1$-minute bins. The central figure magnifies $10$ minutes of the downlink data rates starting at 9:47 PM and averaging the results over $5$ second bins, separating the contributions of the $5$ most active users and aggregating the rest. Finally, the figure on the right plots the variation of the MCS on $2$-second bins, obtained by the second user of the central graph. Error bars are plotted one standard deviation above and below the average measure. 

\section{Conclusions}
\label{sec:conclusions}

In this paper we introduced OWL, an online and reliable solution to decode the LTE control channel and obtain the complete schedule information of the monitored eNodeB. OWL reliability is achieved by exploiting the LTE random access procedure that, in turn, allows our tool to obtain a list of the active RNTIs of UEs connected to the eNodeB.

We evaluated the performance of our solution by checking the actual utilization of the physical downlink shared channel and we found that not only did OWL decode almost all the RBs in the majority of the tests, but also it significantly outperformed the only comparable non-commercial tool greatly reducing the amount of undetected DCIs.

Finally, by means of a full day measurement campaign we showed how our solution can enable a variety of different analyses which can be subsequently used as both a validation tool for mobile phone measurements and for measuring and testing the performance of existing and future LTE deployments.

\vspace{.4cm}
\section*{Acknowledgments}
This work has been supported by the European Union H2020-ICT grant 644399 (MONROE), by the Madrid Regional Government through the TIGRE5-CM program (S2013/ICE-2919), the Ramon y Cajal grant from the Spanish Ministry of Economy and Competitiveness RYC-2012-10788 and grant TEC2014-55713-R.

\vspace{.6cm}
\bibliographystyle{abbrv}
\bibliography{paper}

\end{document}